\makeatletter
\declare@file@substitution{revtex4-1.cls}{revtex4-2.cls}
\makeatother
\documentclass[aps,prd,twocolumn,showpacs,superscriptaddress,groupedaddress,nofootinbib]{revtex4-2}
\usepackage[portuges, english]{babel}
\usepackage[utf8]{inputenc}
\usepackage{amsmath}
\usepackage[makeroom]{cancel}
\usepackage{slashed}
\usepackage{enumitem}
\usepackage{graphicx}
\usepackage[caption=false]{subfig}
\usepackage{natbib}
\usepackage[hidelinks,colorlinks=true,linkcolor=red,citecolor=red]{hyperref}

\usepackage{tikz,hyperref, xcolor}
\definecolor{lime}{HTML}{A6CE39}
\definecolor{Emerald}{HTML}{50c878}
\definecolor{PineGreen}{HTML}{01796F}
\definecolor{ForestGreen}{HTML}{228B22}
\definecolor{Coral}{HTML}{FF7F50}
\definecolor{YellowOrange}{HTML}{E94E16}
\definecolor{DarkBlue}{HTML}{1A237E} 
\definecolor{DarkRed}{HTML}{8B0000}  
\hypersetup{linkcolor=Emerald,citecolor=red,filecolor=red,
urlcolor=blue}

\def\adv{AIP Advances}%
\def\apj{ApJ}                 
\def\apjl{ApJ}               
\def\aap{A\&A}                
\def\jcap{JCAP}              %
\def\mnras{MNRAS}             
\def\prc{Phys.~Rev.~C}        
\def\prd{Phys.~Rev.~D}        
\def\prl{Phys.~Rev.~Lett.}    

\def\nphysa{Nucl.~Phys.~A}   
\def\physrep{Phys.~Rep.}   

\begin{document}

\medskip\noindent

\title{\Large  Rotational Behaviour of Exotic Compact Objects}

 \DeclareRobustCommand{\orcidicon}{%
 	\begin{tikzpicture}
 		\draw[lime, fill=lime] (0,0) 
 		circle [radius=0.16] 
 		node[white] {{\fontfamily{qag}\selectfont \tiny ID}};
 		\draw[white, fill=white] (-0.0625,0.095) 
 		circle [radius=0.007];
 	\end{tikzpicture}
 	\hspace{-2mm}
 }
 \foreach \x in {A, ..., Z}{%
 	\expandafter\xdef\csname orcid\x\endcsname{\noexpand\href{https://orcid.org/\csname orcidauthor\x\endcsname}{\noexpand\orcidicon}}
 }
\def\adv{AIP Advances}%
\newcommand{\orcidauthorB}{0000-0003-4777-4188} 
\author{Zakary Buras-Stubbs\orcidB{}}
\affiliation{Centro de Astrof\'{\i}sica e Gravita\c c\~ao  - CENTRA, \\
	Departamento de F\'{\i}sica, Instituto Superior T\'ecnico - IST,\\
	Universidade de Lisboa - UL, Av. Rovisco Pais 1, 1049-001 Lisboa, Portugal}
\email{zburasstubbs@tecnico.ulisboa.pt}
%
\newcommand{\orcidauthorA}{0000-0002-5011-9195} 
\author{Ilídio Lopes\orcidA{}}
\affiliation{Centro de Astrof\'{\i}sica e Gravita\c c\~ao  - CENTRA, \\
	Departamento de F\'{\i}sica, Instituto Superior T\'ecnico - IST,\\
	Universidade de Lisboa - UL, Av. Rovisco Pais 1, 1049-001 Lisboa, Portugal}
\email{ilidio.lopes@tecnico.ulisboa.pt}




\begin{abstract}
\noindent	
We construct exotic compact objects composed entirely of self-interacting asymmetric fermionic dark matter governed by a repulsive Yukawa potential with massive dark interaction boson. By considering the structural, tidal, and rotational properties of solar mass self-gravitating dark matter systems, and contrasting them against purely baryonic neutron stars, described by the well understood SLy4 equation of state, we hope to shed some light on the place of dark compact systems in the context of gravitational wave astronomy, specifically due to the difficulty parsing mass and radius data from events with no electromagnetic counterpart. Here we consider systems composed of $1\,\mathrm{GeV}$ and $10\,\mathrm{GeV}$ dark matter. Relevant compact objects are then analysed and simulated as both static bodies, and rotating systems governed by the Hartle-Thorne formalism to second order.
\\Here within we highlight the differences in key tidal and rotational properties encoded in gravitational wave signals, and analyse how dark objects may mimic or distinguish themselves to current and future gravitational wave observatories. 

\end{abstract}


\maketitle

\section{Introduction} \label{sec:intro}

\medskip\noindent
Modern astrophysics and cosmology still faces challenges in finding apparent hidden mass in the Universe first proposed almost a century ago \cite{2017arXiv171101693A}, with the first true implications of its existence becoming clear around fifty years later through observations of galactic rotational curves \cite{1980ApJ...238..471R}. 
\\Within this time a veritable zoo of particle physics models have come, gone, and returned in the search for a clean undeniable solution to the origin of a majority of the mass in the Universe \cite{2021PrPNP.11903865A}, and indeed due to how dark matter seems to dominate the Universe it is quite possible that the dark sector exhibits its own rich structure not simply solved by a single particle species. 
\\The most commonly used solution for an extended period came to be that of cold collisionless dark matter (CCDM), as it effectively replicated most of the broader structure observed in the Universe \cite{2021PrPNP.11903865A,2006ApJ...648L.109C}. However as observational techniques rapidly improved it became clear that the collisionless paradigm struggled to explain discrepancies in small scale structural behaviour in the context of galactic core structure and absence of star forming in regions of high expected mass density \cite{2018PhR...730....1T}.
\\These discrepancies have opened the door for an even greater variety of proposals, from considerations of simple, well-understood particle behaviours to hypothetical particles stemming from natural Standard Model extensions\cite{2004astro.ph.12170B}, with no clear resolution without the advent of novel observational techniques.
\\In this vein, we opt to consider the existence of objects entirely composed of accumulated dark matter, focusing on the parameter space that is capable of generating objects that may hide amongst or mimic other solar mass cold compact objects, such as neutron stars.In this way, by considering the structural and tidal properties of a wide parameter space of dark matter systems, we may identify ways in which these objects might reveal themself in merger events.
\\ Due to the potential vastness of models that may combine to form the dark sector as a whole, we opt to proceed cautiously and pragmatically in selecting from the vast dark parameter spaces that are not in conflict with observational constraints, which in turn might offer an opportunity to confirm some subset of models via gravitational wave detection.
\\The rest of this paper deals with the underlying physics involved in constructing and modelling such systems and their ramifications, and is organised as follows. In Sec. \ref{sec:SPAEOS}, we layout the static relativistic equations used, as well as the particle physics and microphysics defining the equation of state for our dark matter systems, and that of the neutron star model that we will use for comparison. We finish this section by highlighting quantities pertaining to tidal deformability in this static case. In Sec. \ref{sec:HTPE}, we underline the relativistic equations describing the rotational modifications at second order made using the Hartle-Thorne formalism, as well as defining further quantities important to measuring structural properties. In Sec. \ref{sec:R}, we analyse and contrast the results of static and rotational object simulations. Finally in Sec. \ref{sec:Con}, we summarise and conclude our findings.
 \\Here, unless specified otherwise, we utilize natural units, where the speed of light ($c$), the gravitational constant ($G$), and the reduced Planck constant ($\hbar$)  are set to $1$.
\section{Static Physics and Equations of state}  
\label{sec:SPAEOS}  
 \medskip\noindent 
We begin by unpacking the basic physics involved in the systems explored in this article, namely the relevant relativistic equations, the microphysics that underpins the dark matter models we aim to investigate, as well as the neutron star EOS with which they will be compared. We summarise how we quantify the tidal deformability of an object.

\subsection{Tolman–Oppenheimer–Volkoff Equations}

 \medskip\noindent 
The action of a simple self-gravitating fluid will necessarily take the following standard form,
\begin{equation} \label{eq: S}
	S = S_G+S_M,
\end{equation}
where $S_G$ is the Einstein-Hilbert action, containing the purely gravitational properties of the system, and $S_M$ is the action concerned with the properties of the self-gravitating perfect fluid constituting the object. Here we consider that our static system is isotropic, meaning that we assume the behaviour of the perfect fluid can be fully defined by its density $\rho$, radial pressure $P$, and an equation of state linking the two. This need not always be the case and one may find an example of an analysis into anisotropic matter in \cite{2024PhRvD.109d3043B} or \cite{2019EPJP..134..454L}.

\medskip\noindent
Einstein's equations may be derived in the usual way, via variation of the total action with respect to the metric, providing
\begin{equation}
\label{eq: EEq}
G_{\mu\nu} = R_{\mu\nu} - \frac{1}{2}g_{\mu\nu}R = 8 \pi T_{\mu\nu}
\end{equation}
where the stress energy tensor can be expressed as,
\begin{equation} \label{eq: Tmn}
	T^\mu_\nu = {\rm Diag}\left(-\rho, P, P, P\right).
\end{equation}

\medskip\noindent
For static space-time we may define the metric to be
\begin{equation} \label{eq:g}
ds^2= -e^\nu dt^2 + e^\lambda dr^2 + r^2d\Omega^2,
\end{equation}
where $\nu (r)$ and $\lambda(r)$ are the metric potentials, and $d\Omega^2 \equiv (d\theta^2 + \sin^2{\theta}\;d\phi^2)$ represents the solid angle element. Einstein's equations may be then used to furnish,
\begin{equation} \label{eq: B}
	e^{\lambda(r)} = \frac{1}{1 - \frac{2m}{r}},
\end{equation}
and
\begin{equation} \label{eq: lam}
	\nu^\prime(r) = \frac{2 m + 8 \pi r^3 P}{r^2 - 2 m r },
\end{equation}
from which we may then derive the generalised Tolman–Oppenheimer–Volkoff (TOV) equation \citep{1939PhRv...55..374O,1939PhRv...55..364T},
\begin{equation} \label{eq:TOV}
	\frac{dP}{dr} = -\frac{(\rho + P)(m + 4\pi r^3 P)}{r(r - 2m)},
\end{equation}
describing the balance of gravitational and hydrostatic forces within a system, where the mass at any point $r$ is described by,
\begin{equation} \label{eq: dmdr}
	\frac{dm}{dr} = 4\pi r^2 \rho.
\end{equation}

\subsection{Equations of State}
\subsubsection{Dark Matter}
\begin{figure}
 \centering{\hspace{-0.8cm}\includegraphics[width=1.1\linewidth]{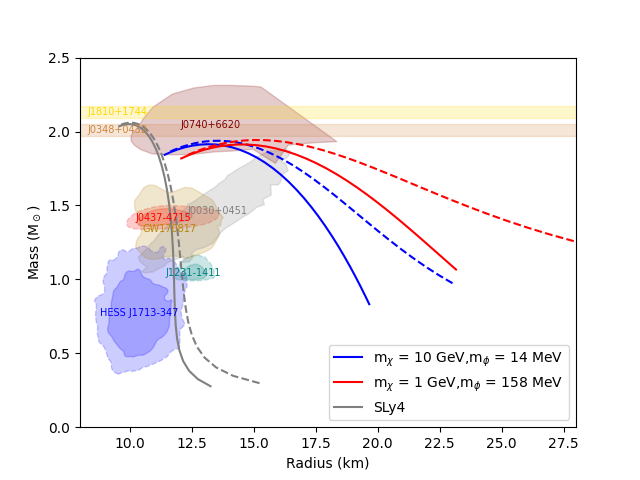}}
 \caption{Mass radius relations of static (solid) and rotating (dashed) compact objects based on the equations of state discussed in Sec. \ref{sec:SPAEOS}. The yellow stripe refers to the 1$\sigma$ constraint on the mass of PSR J1810+1744 \cite{2021ApJ...908L..46R}, and the brown stripe to the 1$\sigma$ constraint on PSR J0348+0432 \cite{2013Sci...340..448A}. The maroon region corresponds to  PSR J0740+6620 \cite{2024ApJ...974..295D}, the gray to PSR J0030+0451 \cite{2019ApJ...887L..24M}, tan to GW170817 \cite{2018PhRvL.121p1101A}, red to J0437-4715 at the $95\%$ confidence level (CL) for the outer region and $68\%$CL inner \cite{2024ApJ...971L..20C}, the teal regions to J1231-1411($68\%$CL inner,$95\%$CL outer)\cite{2024ApJ...976...58S}, and the blue regions to HESS J1713-347($68\%$CL inner,$95\%$CL outer)\cite{2022NatAs...6.1444D}. Rotating objects revolve at a frequency of 400Hz in all cases, and objects composed entirely of dark matter are indicated in red and blue, with the relevant respective masses indicated.}
    \label{fig:MR}
\end{figure}

The particular form that dark matter might take remains highly uncertain, with very few constraints providing researchers with a clear avenue of attack on the problem aside from implied astrophysical mass distributions and limits on interactions cross-sections.
\\The most commonly used choice is that of collisionless cold dark matter (CCDM), however to date this has come with associated astrophysical structural problems that are not reflected in observation, namely the missing satellites, core-cusp, and too-big-to-fail problems, a review of which may be found in \cite{2018PhR...730....1T}.
\\Although these discrepancies may be individually resolved with relative ease by the introduction of self interaction into a dark matter model, restrictions must be placed on these to ensure that the scattering cross-section is sufficiently large to produce cores observed within galaxies, without being so large that it completely disrupts the structure of galactic clusters. Historically, reconciling these two scales with a constant interaction cross-section has proven difficult, as the ratio of the scattering cross section to the particle mass required to flatten dwarf galaxy central densities would typically evaporate elliptical galaxy halos in hot clusters.
\\This discrepancy could potentially be resolved by the consideration that the interaction cross-section is instead velocity dependant, for example as in \cite{2013PhRvD..87k5007T}, as the characteristic velocities in galactic clusters are $\sim100$ times greater than on the scale of dwarf galaxies. This behaviour would also be in keeping with our understanding of particle physics due to a similar deviation from Rutherford scattering at high energies once the momentum transfer becomes similar to the mass of the force carrier.
\\A general review of the topic of reconciling the microscopic scale with astrophysical structure may be found in \cite{2024ARNPS..74..287Z}. 
\\
\\Even with these constraints, there is still very little that might elucidate the fundamental nature of particle dark matter, for instance whether it is fermionic or bosonic, never mind the precise nature of interactions that may exist in the dark sector.
\\As our approach to particle selection in this paper is based on investigating possible exotic objects that may hide amongst ordinary baryonic compact objects, we opt to constrain ourselves to repulsive asymmetric fermionic dark matter, as this is a natural extension mimicking the baryonic matter that we observe in the Universe.
\\
\\Due to the high density environment being replicated in $\sim$M$_\odot$ systems, the skeleton of our fermionic equation of state may be quite reasonably described by a Fermi gas at zero temperature. For a spherical body composed of degenerate Fermi gas, the Fermi momentum may be found to be,
\begin{equation}
    k_F = (3\pi^2n)^{\frac{1}{3}},
\end{equation}
where $n$ is the number density of fermionic particles, and we have assumed a degeneracy of $g=2$.
\\For this system the energy density and pressure as a function of the Fermi momentum are well known and, by considering their statistical distribution, may be found to be,
\begin{align}
    \rho =& \frac{1}{\pi^2}\int_0 ^{k_F} k^2 \sqrt{m_\chi^2 +k^2}dk &\\ \nonumber=& \frac{m_\chi^4}{8\pi^2}\left((2x^3+x)\sqrt{1+x^2} - \sinh^{-1}{(x)}\right)    
\end{align}
\begin{align}
     P =& \frac{1}{3\pi^2} \int_0 ^{k_F} \frac{k^4}{\sqrt{m^2_\chi +k^2}}dk \\\nonumber=& \frac{m_\chi^4}{24 \pi^2} \left((2x^3-3x)\sqrt{1+x^2} + 3\sinh^{-1}{(x)}\right)
\end{align}
where $x = \frac{k_F}{m_\chi}$. 
\\Interactions may then be included in their most basic form by considering only two-body interactions between dark matter particles, $\chi\chi \rightarrow\chi\chi$. Via Hartree-Fock, the lowest order interaction energy density can be expected to be proportional to the particle number density squared, $n^2$. To ensure that this has the correct dimensions as energy density we introduce the mass scale of the interaction, $m_I$, by which we may receive an approximate description of the energy density due to self interaction,
\begin{equation}
    \rho_{I}= \frac{n^2}{m_I^2}.
\end{equation}
\\The fundamental thermodynamic relation then tells us,
\begin{equation}
    P_I = -\frac{\partial E}{\partial V}|_S = n^2 \frac{\partial\rho/n}{\partial n} = \frac{n^2}{m_I^2} = \frac{1}{9\pi^4}\frac{k_F^6 }{m_I^2},
\end{equation}
and hence we find that the interaction term for both the pressure and density of the system is the same.
Incorporating the effects on energy density and pressure created by this approximate inclusion of a self interaction we finally have,
\begin{align} \label{rhoeos}
    \rho =& \frac{m_\chi^4}{9\pi^4}x^6 \lambda^2+\\ \nonumber& \frac{m_\chi^4}{8\pi^2}\left((2x^3+x)\sqrt{1+x^2} - \sinh^{-1}{(x)}\right)    
\end{align}
\begin{align}\label{peos}
     P =& \frac{m_\chi^4}{9\pi^4}x^6 \lambda^2+\\\nonumber& \frac{m_\chi^4}{24 \pi^2} \left((2x^3-3x)\sqrt{1+x^2} + 3\sinh^{-1}{(x)}\right)
\end{align}
where $\lambda = \frac{m_\chi}{m_I}$ is a measure for the strength of the repulsive self-interaction. The equation of state for this system may then be determined parametrically from $x$ for given choices of $m_\chi$ and $m_I$.
\begin{figure}
 \centering{\includegraphics[width=1\linewidth]{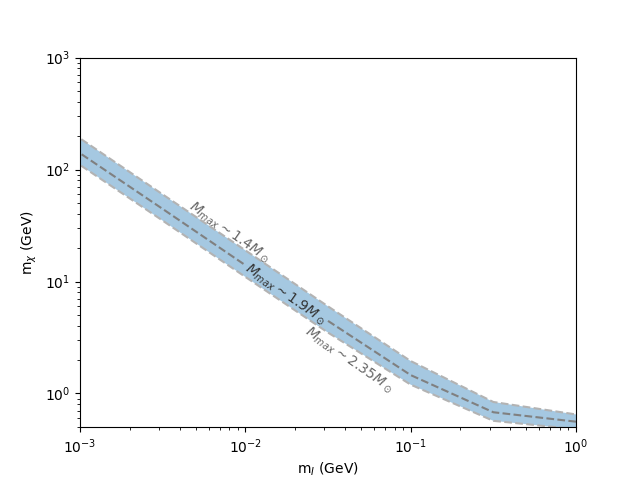}}
 \caption{ Graph of dark fermion mass and dark interaction mass scale, $m_\chi$ and $m_I$. Curves indicate the parameter space of dark fermion and dark interaction boson masses that produce dark compact objects whose maximum mass is similar to the mass of other cold compact objects, for example neutron stars. Dashed lines indicate parameter configurations that produce objects that can attain the same indicated maximum mass across the parameter space. The distinct gradient change present in the curves is most likely determined by which term is a dominant factor in equations \ref{rhoeos} and \ref{peos}, where the leftmost region is dominated by the dark interaction, and rightmost dominated by fermionic nature of the dark matter particles.}
 \label{fig:mchimphi}
\end{figure}
\\
\\As the systems being simulated are those of degenerate Fermi gases, the most comparable baryonic system is that of a neutron star, and hence we opt to focus on objects of similar masses. A graph of the relevant portion of the parameter space to be investigated may be seen in Figure \ref{fig:mchimphi}, where curves indicating the maximum stable mass thresholds observed in simulated objects can be seen. Here, the upper mass limit of $2.35$M$_\odot$ was selected to match that of the most massive neutron star observed to date \cite{2022ApJ...934L..17R}, and the lower limit of $1.4$M$_\odot$ chosen to approximate an average neutron star mass. Objects explored in this paper will have a maximum mass of $1.9$M$_\odot$ to act as a reasonable medium between the two.
\\The validity of this region within a wider context can be verified in several ways. First, within the context of direct analysis of required interaction cross-sections, it is well observed that in dwarf galaxies an interaction cross-section specified by $\sigma/m_\chi \sim 10 \text{cm}^2/\text{g}$ may be required to flatten out the galactic core and avoid a cusp, and similarly for galaxy clusters it is known that the cross-section is limited to be less than $\sigma/m_\chi \sim 0.1 \text{cm}^2/\text{g}$ \cite{2012MNRAS.423.3740V,2013MNRAS.430...81R}.
\\Expressing this in a slightly different way we see that,
\begin{equation}
    \sigma \sim (0.1-10) \cdot(m_\chi/g) \text{cm}^2 \approx 1.8 \cdot 10^{-24} \cdot(0.1-10) \cdot(\frac{m_\chi}{\mathrm{GeV}})\text{cm}^2 ,
\end{equation}
which is notably significantly larger than the weak force cross-section, $\sigma\sim10^{-38}$, and hence implying that the dark interaction mediator must be significantly lighter than the energy scale of the weak force, that is $m_\phi\ll 300\,\mathrm{GeV}$. 
\\The specific model governing our dark matter is that of a Dirac fermion $\chi$ coupled to a massive vector interaction boson $\phi$, with mass scale $m_\phi \sim m_I$, by 
\begin{equation}
    \mathcal{L}_{int} = g_\chi \bar{\chi}\gamma^\mu\chi\phi_\mu,
\end{equation}
where $g_\chi$ is the coupling constant, and $\gamma^\mu$ are the Dirac matrices. The non-relativistic form of this interaction may be modelled by a Yukawa potential,
\begin{equation}
    V = \pm\frac{\alpha_\chi}{r}e^{-m_\phi r}
\end{equation}
for which $\chi\chi \rightarrow\chi\chi$ is repulsive(+), and $\chi\bar{\chi} \rightarrow\chi\bar{\chi}$ is attractive(-).
\\Note that this is not the only choice of model that can produce the equation of state defined by \ref{rhoeos} and \ref{peos}, as the estimate for interaction energy density is made at the lowest possible order. As a result an attractive vector or scalar boson interaction also result in this EOS, as is explored in \cite{2024MNRAS.527.6795M}, however the constraints placed on this style of dark matter are drastically different, and as such the parameter space explored in this paper is for the most part not applicable.
\\Simulations may be performed to determine the parameter space of particles that satisfies the constraints on particle interaction cross-sections on a dwarf galaxy, and cluster scale, examples of which may be seen in \cite{2018PhR...730....1T},\cite{2013PhRvD..87k5007T}, and \cite{2016PhRvL.116d1302K}. 
\\Due to there still being a great deal of freedom in possible choices of the dark fine structure constant, $\alpha_\chi$, the allowed parameter space identified may still be relatively large, and indeed covers a majority of the region identified in \ref{fig:mchimphi} if we approximate the interaction mass scale $m_I$ to be equivalent to $m_\phi$. 
\\As a result the dark fermion masses that will be explored in this paper are $m_\chi = 10\,\mathrm{GeV}$ and $1\,\mathrm{GeV}$, as they provide a reasonable range encompassed by our parameter space of interest in Figure \ref{fig:mchimphi} and allowed regions constrained by astrophysical observations and simulations for repulsive SIDM models \cite{2018PhR...730....1T,2013PhRvL.110k1301T}, as well as spanning a region in which $m_\phi\ll m_\chi$, ensuring that the interaction will be sufficiently long ranged. For completeness, objects composed of more massive dark matter considered as a toy model, for example  $m_\chi = 100\,\mathrm{GeV}$ , were found to differ very little from objects comprised of $10\,\mathrm{GeV}$ particles, likely due to the interaction terms in equations \ref{rhoeos} and \ref{peos} being the dominant factors in both. In this particular instance a choice of $m_\chi = 100\,\mathrm{GeV}$ and $m_\phi = 1.4 \,\mathrm{MeV}$ results in a curve that is indistinguishable from that produced by $m_\chi = 10\,\mathrm{GeV}$ and $m_\phi = 14\,\mathrm{MeV}$ in Figure \ref{fig:MR}.
\subsubsection{Neutron Star}
\label{subsec:NS_EOS}

\medskip\noindent
Neutron star structure is governed by an equation of state (EOS) that links the outer crust, inner crust, and uniform core in a thermodynamically consistent way. Several well-established classes are used in the literature. Non-relativistic microscopic nuclear many-body approaches utilise Skyrme energy-density functionals such as SLy \citep{1997NuPhA.627..710C,2001A&A...380..151D}, BSk \citep{2010PhRvC..82c5804G}, and KDE \citep{2005PhRvC..72a4310A}, fitted to ground-state data of finite nuclei and constrained by neutron-matter calculations. Relativistic mean-field models, for example DD2 \citep{2010PhRvC..81a5803T}, SFHo \citep{2013ApJ...774...17S}, TM1 \citep{1994NuPhA.579..557S}, GM1 \citep{1991PhRvL..67.2414G}, and NL3 \citep{1997PhRvC..55..540L}, encode interactions through meson–nucleon couplings that may be density dependent. Ab initio chiral effective field theory \citep{2013ApJ...773...11H,2018ApJ...860..149T} provides controlled information up to roughly twice nuclear saturation density, and is commonly extended to higher densities using spectral \citep{2018ApJ...860..149T} or piecewise-polytropic \citep{2009PhRvD..79l4032R} parameterisations. Further variants introduce additional degrees of freedom, such as hyperons \citep{2017PhRvC..95f5803F} or deconfined quarks \citep{2018PhRvL.120z1103M}, whilst hybrid constructions allow first-order phase transitions \citep{2019PhRvD..99j3009M,2019PhRvL.122f1102B}. Phenomenological EOS with induced surface tension yield unified descriptions compatible with heavy-star constraints \citep{2017ApJ...850...75S,2019ApJ...871..157S}, and rotational limits from the fastest millisecond pulsars may offer complementary indicators of quark matter \citep{2025PhRvD.111l3021G}. These choices differ in stiffness above saturation, which sets mass–radius relations, compactness, and the tidal response.

\medskip\noindent
Multi–messenger observations now constrain very soft and very stiff EOS. The tidal deformability inferred from GW170817 favours moderate pressures around two to three times saturation density \citep{2018PhRvL.121p1101A}. NICER pulse–profile modelling of PSR~J0030+0451 and PSR~J0740+6620 yields radii near 12 to 14 km for masses from about 1.4 to above 2.0~$M_\odot$, and therefore requires an EOS that supports at least two–solar–mass stars while avoiding excessive radii \citep{2021ApJ...918L..28M,2024ApJ...974..295D}. These bounds are consistent with recent global reviews \citep{2017RvMP...89a5007O}.

\medskip\noindent
In this work, we adopt the unified SLy4 EOS as our baryonic benchmark. The rationale is fourfold. First, SLy4 is unified from the outer crust to the liquid core within a single Skyrme functional, which avoids arbitrary joins and preserves continuity of pressure, chemical potentials, and sound speed \citep{2001A&A...380..151D}. Secondly, the underlying interaction was tuned to neutron–rich systems and exhibits compatibility with many–body constraints on pure neutron matter where chiral EFT remains reliable \citep{1998NuPhA.635..231C}. Thirdly, mass–radius sequences computed with SLy4 lie within the credible regions set by GW170817 and NICER while supporting maximum masses above $2
\,M_\odot$ \citep{2018PhRvL.121p1101A,2021ApJ...918L..28M}. Fourthly, SLy4 provides a widely used crust for thermal evolution and oscillation studies, which facilitates direct comparison with previous results reported in this article \citep{2001A&A...380..151D}.

\medskip\noindent
We emphasise the limits of this benchmark. SLy4 is a cold, nucleonic EOS without explicit hyperons or quarks and does not include finite–temperature effects that are relevant for mergers and supernovae. These caveats do not affect the quasi–static stellar models considered here.
Throughout, SLy4 serves as a reference curve rather than a unique description, and we indicate where our conclusions are insensitive to reasonable EOS variation.

\subsection{Tidal Deformability}
As we have concerned ourselves with objects that have gravitational masses indistinguishable
indistinct from that of neutron stars, we must consider differences in structural properties that may reveal themselves in the context of gravitational waves emitted from their mergers with other compact objects, most relevant of which is their tidal deformability.
\\To investigate these properties of our compact objects we make use of the process laid out in \cite{2008ApJ...677.1216H}, where the relevant equations and operations used are as follows.
\\We integrate the differential equation describing the perturbed metric element $H$,
\begin{align}
    H^{\prime\prime} + H^{\prime}\left(\frac{2}{r} + \frac{1}{1-\frac{2m}{r}}\Big[\frac{2m}{r}+4\pi r(p-\rho)\Big]\right) +
    \\\nonumber H \Bigg(-\frac{6}{r-2m}+\frac{4\pi}{1-\frac{2m}{r}}\Big[5\rho + 9p +(p+\rho)\frac{d\rho}{dp}\Big]\\\nonumber-\Big(\frac{2 m + 8 \pi r^3 P_r}{r^2 - 2 m r }\Big)^2\Bigg) =0
\end{align}
from the centre of the system to its surface. With this we may then determine the tidal Love number
\begin{align} \label{eq:k2}
    k_2=& \frac{8 C^5(1-2C)^2}{5}[2+2C(y-1)-y]\times
    \\&\nonumber \Big(2C(6-3y+3C(5y-8))+
    \\&\nonumber 4C^3(13-11y+C(3y-2)+2C^2(1+y))+
    \\&\nonumber 3(1-2C)^2(2-y+2C(y-1))\log(1-2C)\Big)^{-1}
\end{align}
where $R$ is the radius of the object, $C = \frac{M}{R}$ is its compactness , and \\$y = \frac{RH^\prime(R)}{H(R)}$. The tidal deformability and dimensionless deformability can then be found to be $\lambda = \frac{2}{3}k_2 R^5$ and $\Lambda = \frac{2}{3}k_2C^5$ respectively.

\section{Hartle-Thorne Perturbation Equations} 
\label{sec:HTPE}  
 \medskip\noindent 

Here we go into detail on how the static theory is modified by rotational effects up to second order, as laid out by Hartle and Thorne \cite{1968ApJ...153..807H}. 

\medskip\noindent	
The Hartle-Thorne  formalism introduces perturbation functions that describe the object's deformation due to rotation, including changes in the metric, density, and pressure distributions. The metric is expanded to second order in the object's angular velocity, incorporating effects such as frame-dragging and quadrupole deformation.
This method is particularly effective for objects rotating below their Kepler frequency, typically around 1.0--1.5~kHz for a 1.4~M$_\odot$ neutron star, depending on the equation of state \citep{1994ApJ...424..823C, 2016MNRAS.459..646B}. The Hartle-Thorne equations enable the calculation of important properties such as the object's moment of inertia, quadrupole moment, and ellipticity, which are crucial for understanding structural properties and gravitational wave emission from rotating compact objects and binary systems \citep{2013MNRAS.429.3007P}. By extending this formalism to include exotic objects composed of dark matter, we can investigate how these objects may manifest, and whether significant differences may be observed as compared to well understood compact bodies such as neutron stars, and as a further probe in addition to the static tidal properties.

\medskip\noindent
The Hartle-Thorne metric is given by
\begin{align} \label{eq:HT}
	ds^2= &-e^{\nu}[1+2(h_0 + h_2 P_2)] dt^2 \\ \nonumber 
	&+ \frac{[1+2(m_0+m_2P_2)/(r-2M)]}{1-2M/r}dr^2\\
	&+ r^2[1+2(v_2-h_2P_2)][d\theta^2+\sin^2\theta(d\phi-\omega dt)^2],\nonumber
\end{align}
where $h_0$, $h_2$, $m_0$, $m_2$, $v_2$ are perturbation functions of the order $\Omega^2$, where $\Omega$ is the body's angular velocity relative to a distant observer, and $P_2 = (3\cos^2\theta -1)/2$. 

\medskip\noindent
The pressure and energy density of the system differ from the static case in the following way:
\begin{equation} \label{eq: Prot}
	P_{\text{rot}} = P_{\text{sta}} + (\rho_{\text{sta}} + P_{\text{sta}})(p_0 +p_2 P_2)
\end{equation}
\begin{equation} \label{eq: rhorot}
	\rho_{\text{rot}} = \rho_{\text{sta}} + (\rho_{\text{sta}} + P_{\text{sta}})\frac{d\rho_{\text{sta}}}{dP_{\text{sta}}}(p_0 +p_2 P_2)
\end{equation}

\medskip\noindent
The subscript 'sta' indicates physical properties determined by the non-rotating case, and $p_0(r)$ and $p_2(r)$ are the dimensionless pressure perturbation factors determined using the differential equations laid out in the following sections at each point in  the object. These are proportional to $\Omega^2$ and associated with the $l=0$ and $l=2$ spherical harmonics respectively. For the remainder of this work, we choose to omit the subscript 'sta' on properties associated with the static case.

\medskip\noindent
As with most concepts in relativity, the forces experienced by the fluid composing the object will not be determined by $\Omega$, but instead by its angular velocity relative to a local inertial frame. This is done by considering a frame which has freely fallen from infinity to a point $x(r,\theta)$. At this radius the frame will have acquired an angular velocity $\omega$, hence by comparing $\Omega$ with observations made in this frame we can find the relevant angular velocity relative to the local inertial frame denoted by $\bar{\omega} = \Omega - \omega$. This is found by solving:
\begin{equation} \label{eq:OMGEQ}
	\frac{d^2 \bar{\omega} }{dr^2} = -(\frac{4}{r} + \frac{1}{j} \frac{dj}{dr})\frac{d\bar{\omega}}{dr} - \frac{4}{j r} \frac{dj}{dr}\bar{\omega}
\end{equation}
where,
\begin{equation}
	j = e^{-\frac{\nu}{2}}\left(1-\frac{2M}{r}\right)^{\frac{1}{2}}.
\end{equation}
\\Outside of the object one demands that:
\begin{equation}
	\bar{\omega}(r) = \Omega - \frac{2 J}{r^3}
\end{equation}
for a system with total angular momentum $J$. Solving equation \eqref{eq:OMGEQ} as an initial value problem for which $\bar{\omega}(0) = \bar{\omega}_0$ may be chosen arbitrarily and $\frac{d\bar{\omega} }{dr}(0)=0 $. For a body of radius $R$, one may determine the angular momentum of the object as:
\begin{equation}
	J = \frac{R^4}{6}\frac{d\bar{\omega} }{dr}(R)
\end{equation}
and therefore also find the angular velocity seen by a distant observer produced by these initial conditions:
\begin{equation}
	\Omega_0 = \bar{\omega}(R) + \frac{2J}{R^3}.
\end{equation}
Once this is found, and recalling that $\bar{\omega} \propto \Omega$, one may calculate a new choice of $\Omega$ simply by rescaling $\bar{\omega}(r)$, that is:
\begin{equation}
	\bar{\omega}_1(r) = \bar{\omega}_0(r) \frac{\Omega_1}{\Omega_0} 
\end{equation}

\subsection{Metric Perturbation Functions}

\medskip\noindent	
Here,  we analyze the metric functions in the Hartle-Thorne formalism that describe structural deformations of compact objects, focusing on both spherically symmetric ($l=0$) and quadrupolar ($l=2$) components \citep{1968ApJ...153..807H}.

\subsubsection{Spherical Deformations, $l=0$}

\medskip\noindent	
We analyze the stellar structure by solving the spherically symmetric ($l=0$) hydrostatic equilibrium equations, which yield two key quantities: the mass perturbation factor $m_0$ and the dimensionless pressure perturbation factor $p_0$. The governing equations are:

\begin{align} \label{eq:m0eq}
	\frac{dm_0}{dr} = &4\pi r^2 \frac{d\rho}{dP}(\rho + P)p_0 + \frac{j^2 r^4}{12}\left(\frac{d\bar{\omega} }{dr}\right)^2 \\ \nonumber
	&- \frac{2 j r^3 \bar{\omega}^2 }{3} \frac{dj}{dr},
\end{align}
and
\begin{align} \label{eq:p0eq}
	\frac{dp_0}{dr} = &- \frac{m_0(1+8\pi r^2 P)}{(r-2M)^2} - \frac{4\pi(\rho+P)r^2}{(r-2M)}p_0\\
	&+ \frac{r^4 j^2}{12(r-2M)}\left(\frac{d\bar{\omega} }{dr} \right)^2 + \frac{1}{3}\frac{d}{dr}\left(\frac{r^3j^2\bar{\omega}^2 }{r-2M} \right).\nonumber
\end{align}

\medskip\noindent	
The pressure perturbation factor $p_0$ directly influences the rotational corrections to pressure and density through equations \eqref{eq: Prot} and \eqref{eq: rhorot}. Meanwhile, the mass perturbation factor $m_0$ primarily contributes to the total mass of a rotating body with static radius $R$, given by:
\begin{equation}
M_{\text{rot}}(R) = M(R) + m_0(R) + \frac{J^2}{R^3}.
\label{eq:Mrot}
\end{equation}

\subsubsection{Quadrupole Deformations, $l=2$}

\medskip\noindent	
The quadrupolar ($l=2$) deformations of a slowly rotating compact object are governed by a set of coupled differential equations. The metric perturbation functions $v_2$ and $h_2$ characterize the system's response to rotation, determining both its shape deformation and external gravitational field. These functions are crucial for calculating observable properties such as the system's quadrupole moment and tidal deformability - parameters that can be measured through gravitational wave signals from binary neutron star mergers. The governing equations take the form:
\begin{align}
	\frac{dv_2}{dr} = - \frac{d\nu}{dr}h_2 + &\left(\frac{1}{r}+\frac{1}{2}\frac{d\nu}{dr}\right)\\
	&\left[-\frac{2j r^3 \bar{\omega}^2}{3}\frac{dj}{dr} + \frac{j^2r^4}{6}\left(\frac{d\bar{\omega}}{dr}\right)^2\right] \nonumber
\end{align}
\begin{align}
	\frac{dh_2}{dr}= &\left( -\frac{d\nu}{dr} + \frac{r}{\frac{d\nu}{dr}(r-2M)} \left[8\pi (\rho+P) - \frac{4M}{r^3}\right] \right)h_2\nonumber\\
	& -\frac{4v_2}{\frac{d\nu}{dr}r(r-2M)}\nonumber\\
	&+\left[\frac{r}{2}\frac{d\nu}{dr} - \frac{1}{\frac{d\nu}{dr}(r-2M)}\right]\frac{r^3 j^2}{6}\left(\frac{d\bar{\omega}}{dr}\right)^2\\
	& -\left[\frac{r}{2}\frac{d\nu}{dr} + \frac{1}{\frac{d\nu}{dr}(r-2M)}\right] \frac{2jr^2\bar{\omega}^2}{3}\frac{dj}{dr}\nonumber
\end{align}
by which we are finally furnished with the second pressure perturbation factor,
\begin{equation}
    p_2  = -h_2-\frac{1}{3}r^2e^{-\nu}\omega^2.
\end{equation}

\medskip\noindent	
The quadrupole moment $Q$ of a rotating compact object can be calculated by solving the coupled differential equations for the metric perturbation functions $v_2$ and $h_2$, where the exterior spacetime solutions are characterised by:
\begin{align}
h_{2ex} &= KQ_2^2(\zeta) + J^2\left(\frac{1}{Mr^3} + \frac{1}{r^4}\right)\\
v_{2ex} &= \frac{2KM}{\sqrt{r(r-2M)}}Q_1^2(\zeta) - \frac{J^2}{r^4}
\end{align}
where $Q_1^2(\zeta)$ and $Q_2^2(\zeta)$ are associated Legendre functions, $K$ is related to the quadrupole moment, $J$ is the angular momentum, and $\zeta = r/M - 1$.
\\The quadrupole moment $Q$ \citep{1967ApJ...150.1005H}
 for a slowly rotating compact object can be expressed as:
\begin{equation}
Q = \frac{J^2}{M} + \frac{8}{5}\,KM^3.
\label{eq:Qrot}
\end{equation}
for which a dimensionless variant may be expressed as 
\begin{equation}
    \tilde{Q} = \frac{QM}{J^2}.
\label{eq:Qtilde}
\end{equation}
\\In addition to the above we use the eccentricity, $e$ of a rotating object as a further measure of its deformation and the effects of rotation, this is defined as follows,
\begin{equation}
    e= \sqrt{1-\frac{a^2}{b^2}}
\label{eq:ecc}
\end{equation}
where $a$ and $b$ are the rotational axis and equatorial radii of the spheroid respectively.


\section{Results}
\label{sec:R}

 \medskip\noindent 
Here we analyse the results of compact objects modelled using the aforementioned mechanisms. The mass-radius relations of dark matter parametrised by; $m_\chi = 10\,\mathrm{GeV}$, $m_\phi = 14\,\mathrm{MeV}$; and $m_\chi = 1\,\mathrm{GeV}$, $m_\phi = 158\,\mathrm{MeV}$; as well as a neutron star model with the SLy4 EOS to act as a comparison, may be found in Figure \ref{fig:MR}.
\\
\begin{figure}
\centering{\hspace{-0.8cm}\includegraphics[width=1.1\linewidth]{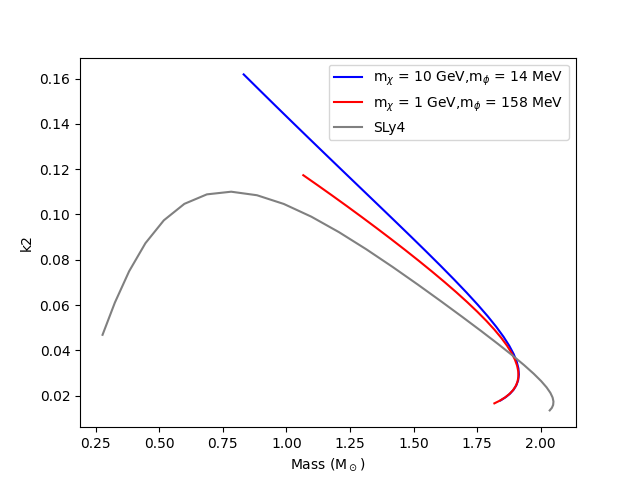}}
\caption{Graph of dimensionless $k_2$ love number against object mass for static systems in Figure \ref{fig:MR}.}
\label{fig:k2}
\end{figure}
\begin{figure}
\centering{\hspace{-0.8cm}\includegraphics[width=1.1\linewidth]{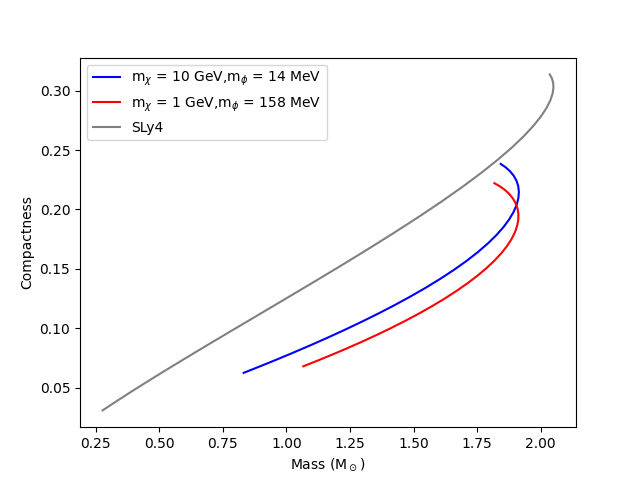}}
\caption{Compactness-mass curves for static objects in Figure \ref{fig:MR}.}
\label{fig:C}
\end{figure}
\begin{figure}
    \centering{\hspace{-0.8cm}\includegraphics[width=1.1\linewidth]{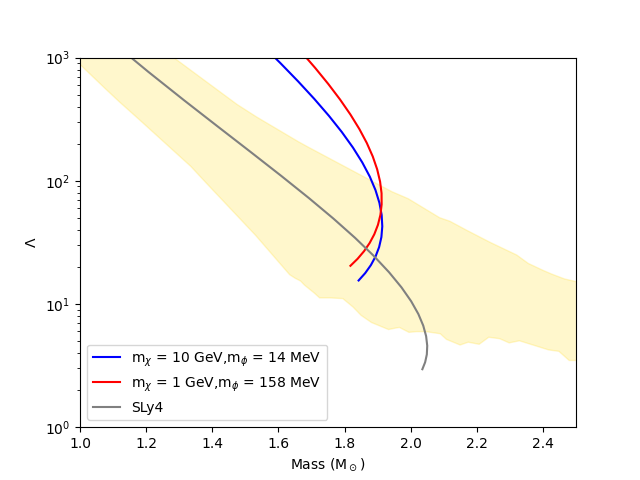}}
    \caption{Graph of the dimensionless tidal deformability number $\Lambda$ against Mass. Shaded in gold is the EOS insensitive tidal deformability constraints placed by GW170817 \cite{2018PhRvL.121p1101A} without a maximum mass constraint. The NS benchmark strongly adheres to observation at all relevant masses, whereas the dark matter objects only fit within these constraints for the highest mass systems. }
    \label{fig:LM}
\end{figure}

\subsection{Static Physics}
\label{sec:R-ASP}

\medskip\noindent
In this section we shall analyse  properties of our static exotic systems as compared to that of a neutron star benchmark. The static objects of interest may be found depicted by solid lines in the mass-radius relation present in Figure \ref{fig:MR}.  For the baryonic benchmark, we adopt the unified SLy4 equation of state (EOS), which combines a microscopic crust with a Skyrme functional core and is thermodynamically consistent across the crust–core transition \citep{2001A&A...380..151D,1997NuPhA.627..710C}. SLy4 describes cold catalysed matter in beta equilibrium, includes leptons, and satisfies standard microphysical requirements of stability and causality in the density range relevant for typical neutron stars.

\medskip\noindent
In our TOV integrations, SLy4 yields mass–radius curves that are compatible with current astrophysical constraints. The radius of a $1.4\,M_\odot$ star lies near the 11.5--12.5 km band commonly inferred from gravitational–wave tidal measurements of GW170817 and from X-ray pulse-profile modelling, while the maximum mass reaches $2.0$--$2.1\,M_\odot$ \citep{2018PhRvL.121p1101A,2024ApJ...974..295D}. This places SLy4 within the allowed region set by the heaviest precisely weighed pulsars and updated NICER analyses of PSR~J0740+6620, which indicate radii around $13\,$km at $2.08\,M_\odot$ and favour slightly softer core pressures than earlier estimates \citep{2024ApJ...974..295D}. Consistency with these data is sufficient for our purpose of contrasting baryonic stars with the self–interacting dark matter models.
\medskip\noindent
The SLy4 curve in Fig. \ref{fig:MR} anchors the baryonic trend in the mass--radius plane, against which the dark--matter sequences are compared. The corresponding compactness $C=M/R$ and tidal Love number $k_2$ follow the expected behaviour: SLy4 stars are systematically more compact at fixed mass and exhibit smaller $k_2$ and $\Lambda=(2/3)\,k_2\,C^{-5}$ than the dark--matter configurations of equal mass, except near the highest central densities where the sequences converge \citep{2008ApJ...677.1216H}. This is seen in Figs.~\ref{fig:k2}, \ref{fig:C}, \ref{fig:LM}, \ref{fig:Ecc}, and \ref{fig:Q}, where SLy4 tracks yield lower deformabilities at given $M$ and $R$, consistent with the higher compactness shown in Fig.~\ref{fig:C}. These patterns agree with the standard scaling of $k_2$ with EOS stiffness and compactness, and with bounds inferred from GW170817 \citep{2018PhRvL.121p1101A} in Fig. \ref{fig:LM}.

\medskip\noindent
As shown in Figure~\ref{fig:MR}, the dark fermion stars with $m_\chi=10\,\mathrm{GeV}$ and $1\,\mathrm{GeV}$ occupy larger radii than SLy4 at a given mass and intersect the observationally allowed bands only near their smallest radii and highest masses. In that regime a self--interacting fermionic compact star can mimic a heavy neutron star in the mass--radius plane, providing an alternative interpretation for very high mass compact objects \citep{2022ApJ...934L..17R,2024ApJ...974..295D,2019ApJ...887L..24M}. The multi--messenger record suggests that a bright electromagnetic counterpart, as observed for GW170817, favours baryonic matter, whereas mergers without a detected counterpart would be consistent with either unfavourable geometry or with binaries involving non--baryonic compact objects \citep{2017PhRvL.119p1101A,2018PhRvL.121p1101A}. We therefore flag the high--mass, low--radius end of the dark--matter sequences as the region where discrimination by tidal parameters and the presence or absence of an electromagnetic counterpart becomes most informative.

\medskip\noindent
Within the dark--matter models, both configurations correlate with data only at the highest masses and lowest radii \citep{2024ApJ...974..295D,2019ApJ...887L..24M}. All sets retain significantly larger radii than the neutron--star baseline, with systems featuring $m_\chi\gtrsim10\,\mathrm{GeV}$ producing the smallest radii among the dark sequences. Since parameters in this range generate very similar structures, we focus on the $m_\chi=10\,\mathrm{GeV}$ branch henceforth.

\medskip\noindent
Despite being the more compact of the two dark sequences, the heavier variant generally exhibits a larger deformability than the $m_\chi=1\,\mathrm{GeV}$ system (Figs.~\ref{fig:k2} and \ref{fig:C}), with $k_2$ converging at the highest masses. At convergence both exhibit very large central densities, where the interaction term in Eqs.~\eqref{rhoeos} and \eqref{peos} scales strongly with number density. In this limit both fluids tend to an effectively linear equation of state,
\begin{equation}
	P \propto \rho,
\end{equation}
and thus share similar fluid properties encoded in $k_2$.

\medskip\noindent
By contrast, the clear separation from the tidal deformabilities of neutron--star models at low masses likely correlates with the presence of a stiff crust in the baryonic case, whereas the Fermi gases diffuse smoothly to zero pressure and density. Overall, SLy4 remains a conservative and widely used neutron--star baseline that satisfies current global constraints and enables a clean structural contrast with the exotic models explored here \citep{2001A&A...380..151D,1997NuPhA.627..710C}.

\medskip\noindent

\subsection{Rotational Physics}
\label{sec:R-RP}

\medskip\noindent
We now analyse the rotating configurations in Fig.~\ref{fig:MR} and contrast them with standard neutron star behaviour. All rotating models are computed in the slow--rotation regime of the Hartle--Thorne formalism\footnote{See the figure captions for the adopted spin and the discussion in Sec.~\ref{sec:HTPE} on the slow--rotation framework.} and assume rigid rotation at a frequency of 400~Hz, a representative value for fast millisecond pulsars and safely below typical Kepler limits; the second--order expansion is therefore applicable \citep{1968ApJ...153..807H}.

\medskip\noindent
The \emph{eccentricity} \(e\) provides a direct, dimensionless measure of spin--induced flattening of a spheroid [eq.~\eqref{eq:ecc}], hence traces the rotational deformability of the object itself. The \emph{spin--induced quadrupole} \(Q\) and its dimensionless counterpart \(\tilde{Q}\equiv QM/J^{2}\) [eq.~\eqref{eq:Qtilde}] characterise how rotation redistributes mass and imprints the exterior field. While distinct from the tidal Love number \(k_{2}\), the spin quadrupole and \(k_{2}\) are known to obey approximate EOS--insensitive I--Love--Q relations, which we use here as a qualitative organising principle \citep{2013PhRvD..88b3009Y,2013MNRAS.429.3007P,2008ApJ...677.1216H}.

\begin{figure}
    \centering{\hspace{-0.8cm}\includegraphics[width=1.1\linewidth]{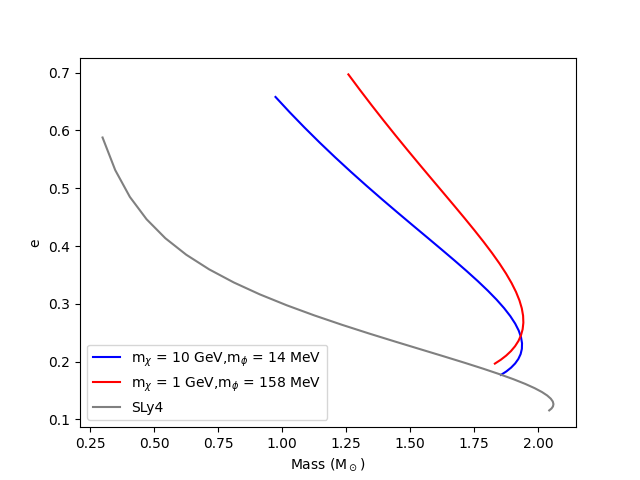}}
    \caption{Graph depicting the eccentricity, $e$, of all rotating objects in Figure \ref{fig:MR} against total mass of each object.}
    \label{fig:Ecc}
\end{figure}

\begin{figure}
    \centering{\hspace{-0.8cm}\includegraphics[width=1.1\linewidth]{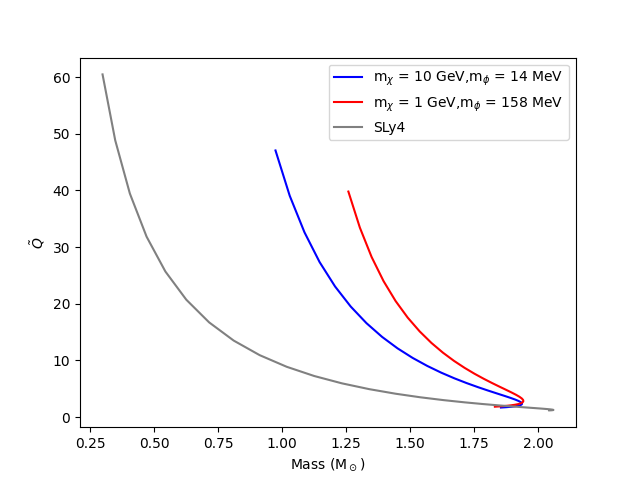}}
    \caption{Graph of dimensionless quadrupole moment, $\tilde{Q}$, against total object mass, for all rotating objects depicted in Figure \ref{fig:MR}.}
    \label{fig:Q}
\end{figure}

\medskip\noindent
These trends are visible across our sequences. Figures~\ref{fig:C} and \ref{fig:Ecc} show that lower compactness correlates with larger rotational eccentricity: more weakly bound configurations are more easily flattened at fixed spin. Conversely, more centrally condensed objects resist rotational deformation and exhibit smaller \(e\). Likewise, Figs.~\ref{fig:k2} and \ref{fig:Q} reveal a progressive convergence of the dark--matter branches at high central density, reflected by similar \(k_{2}\) and \(\tilde{Q}\). This is consistent with the approach to an effectively linear equation of state \(P\propto\rho\) at large number density, which reduces EOS, dependent spread in the rotational response noted in Sec.~\ref{sec:R-ASP}. 

\medskip\noindent
Operationally, our use of rigid rotation implies a single angular velocity throughout the fluid in the local inertial frame, as assumed in the Hartle--Thorne treatment adopted here \citep{1968ApJ...153..807H}. The qualitative behaviours described above agree with the curves displayed for \(e(M)\) and \(\tilde{Q}(M)\) in our models.

 \medskip\noindent
Third-generation ground-based interferometers such as the Einstein Telescope (ET) will observe large samples of binary neutron stars with high signal-to-noise across the low to kHz band, yielding precise measurements of tidal parameters that map directly to \(k_2\) and \(\Lambda = \tfrac{2}{3}k_2 C^{-5}\) \citep{2008ApJ...677.1216H}, and enabling tests of the I-Love-Q systematics of the sequences computed here \citep{2013PhRvD..88b3009Y, 2013Sci...341..365Y} with third-generation detectors \citep{2010CQGra..27s4002P, 2011CQGra..28i4013H, 2020JCAP...03..050M}.
ET is also expected to access the post-merger spectrum of nearby events, where the dominant peak frequency correlates tightly with radii and compactness, providing a complementary route to constrain the equation of state and to search for departures induced by exotic composition \citep{2012PhRvL.108a1101B, 2017ApJ...850L..34B}.  For dark matter admixed stars, ET quality inspirals would test for systematic shifts in \(\Lambda(M)\) relative to hadronic expectations; current forecast studies indicate that robust detection of a percent–level dark matter fraction will require population analyses and joint networks with Cosmic Explorer \citep{2024PhRvD.110j3033K,2024PhRvD.110d3013W}. These capabilities motivate the comparison of \(e\), \(k_{2}\) and \(\tilde{Q}\) presented above with gravitational wave measurable quantities.


\subsection{Combined Structural Properties}

\begin{figure}
\centering{\hspace{-0.8cm}\includegraphics[width=1.1\linewidth]{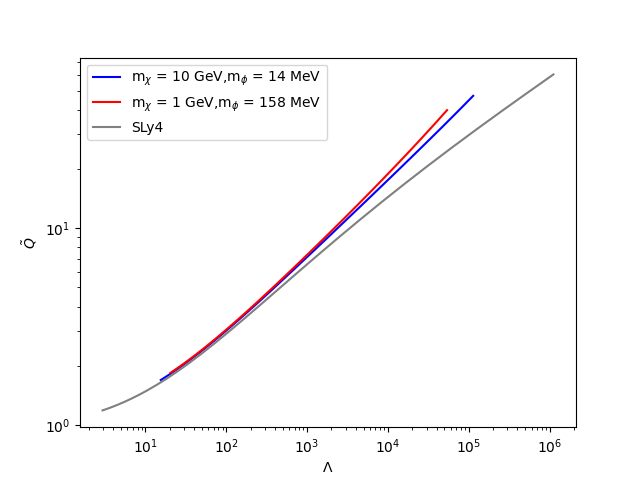}}
\caption{Dimensionless spin-induced quadrupole moment $\tilde{Q}\equiv Q M/J^{2}$ versus the dimensionless tidal deformability $\Lambda$ for rotating compact objects at $400$\,Hz. Curves correspond to dark-matter models with $m_\chi=10$\,GeV, $m_\phi=14$\,MeV (blue) and $m_\chi=1$\,GeV, $m_\phi=158$\,MeV (red), together with the SLy4 neutron-star benchmark (grey). Steeper slopes, $d\tilde{Q}/d\Lambda$, indicate a stronger joint susceptibility to rotational and tidal deformation, reflecting internal-structure differences between exotic and baryonic configurations. The correlations follow the I--Love--Q relations. The $\tilde{Q}$--$\Lambda$ tracks were computed for all objects shown in Fig.~\ref{fig:MR}.} \label{fig:QL}
\end{figure}

\medskip\noindent
By combining static and rotational diagnostics we obtain a clearer view of each model's structure and its imprint on gravitational-wave observables. The response to an external quadrupolar field is quantified by the tidal Love number $k_2$ and the associated dimensionless tidal deformability $\Lambda = \tfrac{2}{3}\,k_2\,C^{-5}$ with compactness $C \equiv M/R$ \citep{2008PhRvD..77b1502F}. 
For baryonic neutron stars, canonical $1.4\,M_\odot$ models give $\Lambda\sim70$--$800$ depending on the stiffness of the equation of state \citep{2008ApJ...677.1216H}. Exotic compact objects, such as self-interacting fermionic dark-matter configurations, can depart systematically owing to distinct pressure--density relations and stratification.

\medskip\noindent
Near-universal I--Love--Q relations link the moment of inertia $I$, the tidal sector via $k_{2}$ (or $\Lambda$), and the spin-induced quadrupole $\tilde{Q}\equiv Q M/J^{2}$ across broad EOS families \citep{2013PhRvD..88b3009Y}. These correlations typically hold to within a few per cent for realistic stellar models and are dominated by outer-layer physics where different EOS tend to converge. Hence, a measurement of any one of $\{I,\Lambda,\tilde{Q}\}$ allows the others to be inferred with minimal microphysical assumptions, while significant deviations from the standard neutron-star track would point to alternative internal physics \citep{2013PhRvD..88b3009Y}.

\medskip\noindent
Third-generation interferometers such as the Einstein Telescope will sharpen constraints on $\Lambda(M)$ and multipole structure through improved low-frequency sensitivity and higher signal-to-noise inspirals \citep{2010CQGra..27s4002P,2011CQGra..28i4013H,2020JCAP...03..050M}.
Through improved noise suppression, the Einstein Telescope targets an order-of-magnitude sensitivity improvement over current detectors across a band spanning a few hertz to several kilohertz \citep{2010CQGra..27s4002P}. Forecasts indicate that well-localised events could enable per-cent level constraints on $\Lambda$ and population studies that probe EOS systematics and the demographics of exotic versus baryonic configurations \citep{2020JCAP...03..050M,2024PhRvD.110d3013W}. Dedicated studies further show that dark-matter effects in neutron-star mergers imprint measurable tidal signatures in the ET band \citep{2024PhRvD.110j3033K}.

\medskip\noindent
In Fig.~\ref{fig:QL}, a steeper track means that small variations in $k_{2}$, and thus in $\Lambda$, correspond to comparatively larger changes in $\tilde{Q}$. The slope $d\tilde{Q}/d\Lambda$ is therefore a useful correlation diagnostic of joint tidal and rotational responses: larger slopes are associated with configurations that are more easily reshaped by an external field, for example those with softer effective pressure support or less centrally concentrated mass profiles. The trends along our sequences follow the I--Love--Q systematics \citep{2013PhRvD..88b3009Y}. Correlating measurements of $\Lambda$ and $\tilde{Q}$ in ET-quality data can therefore provide complementary tests of composition and the high-density equation of state.

 
 \section{Summary and Conclusion} 
\label{sec:Con}

\medskip\noindent

Within this article we have explored the implications of dark matter consistent with astrophysical observations of interaction cross-sections, chosen as a simple yet natural continuation of Standard Model properties, in the context of its accumulation into compact objects, for which the particularities of model selection may be found in Section \ref{sec:SPAEOS}.
\\To do this we considered simulated compact systems through the lens of the conflux of their static, Sec. \ref{sec:SPAEOS}, and rotational properties, Sec. \ref{sec:HTPE}, as well as in comparison to those found in neutron stars described by the SLy4 equation of state, chosen as a baseline due to its efficacy in reproducing canonical behaviours and features of observed neutron stars. By considering compact objects in this way, we hoped to provide ample leeway for analysis and comparison with both contemporary and future observations.
\\The primary dark matter masses explored were $10\,\mathrm{GeV}$ and $1\,\mathrm{GeV}$, with it being noted that the equation of state is such that higher mass particles essentially reproduced interaction driven structures almost identical to the $10\,\mathrm{GeV}$ cases, with an effectively linear equation of state. Similar behaviour may also be observed in EOS sensitive quantities, such as the Love number $k_2$, an indicator of tidal deformability, where both models converge at high masses and hence high densities. In general both models displayed a softer equation of state than the baryonic baseline, with $10\,\mathrm{GeV} $appearing to be the softest but more compact and less rotationally deformable than $1\,\mathrm{GeV}$ due to structural differences between self-gravitating objects produced.  
\\Static properties of each model were contrasted against their rotational behaviour, most notably their spin-induced quadrupole moment against tidal deformability, two parts of the I-Love-Q relations. These indicators are extremely pertinent to current and future gravitational wave observatory missions, as their information is encoded in gravitational wave data and as such offer an opportunity to search for exotic objects for which there is no visible counterpart.
\\The models explored in this paper were designed upon only the most basic principles regarding their microphysics and how that would further manifest in an astrophysical object. We therefore expect that there is ample room to explore this particular model further for the development of more complex systems and processes, as well as to investigate fully distinct models through the same lens.
\\One should note however that within our investigation we did not consider the implications of dark matter choices in the context of the interactions between a dark matter medium and astrophysical bodies contained with in it. For example, in \cite{2025PhLB..86239358S} the authors explore the impact of a dark environment on gravitational wave emission from mergers. Although this is beyond the scope of this paper, the implications of potential Coulomb damping or frequency shifts stemming from specific dark matter choices should not be overlooked
\\Delving into potential compact systems formed by matter that makes up an overwhelming majority of the mass in the Universe offers a fantastic opportunity to rule out or confirm regions of the dark matter parameter space that would otherwise be difficult to identify. This coupled with the increased sensitivity of coming gravitational wave detectors, such as the Einstein Telescope, herald unprecedented possibilities in our continued effort to understand the cosmos.

\section*{Acknowledgments}
\medskip\noindent

I.L. and Z.B.S. gratefully acknowledge the Fundação para a Ciência e Tecnologia (FCT), Portugal, for the financial support to the Center for Astrophysics and Gravitation (CENTRA/IST/ULisboa) through grant No. UID/00099/2025. Z.B.S also acknowledges support provided via a Research
Fellowship contract with Fundação para a Ciência e a Tecnologia  with contract No. BD/02099/2024.
\newpage








 
\end{document}